\documentclass[epj]{svjour}
\usepackage{graphicx,latexsym,color}

\begin{document}

\title{Transport of a quantum degenerate heteronuclear \\
Bose-Fermi mixture in a harmonic trap}

\author{C.~ Klempt, T.~Henninger, O.~Topic, J.~Will, St.~Falke, W.~Ertmer, and J.~Arlt}
\institute{Institut f{\"u}r Quantenoptik, Leibniz Universit{\"a}t
Hannover, Welfengarten 1, D-30167 Hannover, Germany}

\date{\today}

\abstract{We report on the simultaneous transport of mixed quantum
degenerate gases of bosonic $^{87}{\rm Rb}$ and fermionic $^{40}{\rm
K}$ in a harmonic potential. The samples are transported over a
distance of $6~{\rm mm}$ to the geometric center of a
Ioffe-Pritchard type magnetic trap. This transport mechanism was
implemented by modification of the QUIC trap and is free of losses
and heating. It significantly extends the capabilities of this trap
design. We demonstrate a launching mechanism for quantum degenerate
samples and show that highly homogeneous magnetic fields can be
created in the center of the QUIC trap. The transport mechanism may
also be cascaded to cover even larger distances for interferometric
experiments with quantum degenerate samples. 
}
\authorrunning{Klempt et al.}

\maketitle
\section{Introduction}

Within the past decade, the field of ultracold atomic gases has
significantly extended the scope of atomic and molecular
physics~\cite{Anderson1995,Davis1995,DeMarco1999}. The experimental
manipulation of quantum degenerate gases has led to the development
of a toolbox for quantum atom optics \cite{Rolston2002} including
guides \cite{Leanhardt2002}, beam splitters and combiners
\cite{Cassettari2000} as well as switches \cite{Muller2001}. These
tools are intended for the development of a new generation of guided
interferometric sensors
\cite{Hansel2001,Schumm2005,Wang2005a,Fattori2007,Jo2007,Jo2007a,Billy2007}.
One of the main requirements for such experiments is the coherent
spatial transport of quantum degenerate gases. At present,
interferometric applications involving ultracold fermionic atoms are
also evaluated and, in some cases, regarded as superior compared to
their bosonic counterpart
\cite{Andersson2002,Search2003,Modugno2004}. However, adiabatic
transport of quantum degenerate fermionic samples had not been
accomplished up to now.

Three main types of mechanical transport mechanisms for
Bose-Einstein condensates over macroscopic distances have been
reported. The first transport was achieved by carefully moving the
focus of a red-detuned dipole trap over a distance $44~{\rm cm}$
\cite{Gustavson2001}. Although the atomic cloud was heated due to
vibrations of the moving optical components, the finite depth of the
dipole trap provided continuous evaporative cooling. Thus quantum
degeneracy was maintained during the transfer time of $7.5$ seconds.
An alternative method transports the condensate within a
one-dimensional optical lattice \cite{Schmid2006}. In order to
obtain long transport distances, the lattice was produced by two
counter-propagating Bessel beams. By adjusting the relative phase,
it was possible to shift the lattice sites and transport a
condensate over up to $10~{\rm cm}$. The third method involves
lithographic conducting structures, so-called atom chips. After
creating a Bose-Einstein condensate with such a wire structure, it
is possible to shift the condensate by applying modulated currents
to an additional periodical wire pattern
\cite{Haensel2001,Hommelhoff2005,Fortagh2007}. Transfer distances of
up to $1.6~{\rm cm}$ were demonstrated.

However, all of these methods suffer from strong heating or loss
mechanisms. A moving dipole trap involves the translation of optics
and induces heating due to vibration. Optical lattices and chip
traps typically produce strongly confining traps with high trapping
frequencies and high atomic densities. Although high densities
support fast evaporation to quantum degeneracy, they are impractical
for the transport because of heating and atom loss due to enhanced
three-body collision rates.

In this paper, we report on the transport of quantum degenerate
samples of bosonic $^{87}{\rm Rb}$ and fermionic $^{40}{\rm K}$ in a
harmonic potential over a distance of $6~{\rm mm}$. This also constitutes
the first transport experiment with a quantum degenerate Fermi gas. The transport is
realized by adiabatically transforming a Ioffe-Pritchard type
magnetic trap produced by macroscopic coils. The atom numbers are
considerably larger than in experiments with atom chips
\cite{Fortagh2007} and the trapping strength is adjustable without
changing the trap depth.

The initial production of a quantum degenerate Bose-Fermi mixture is
performed in a so-called QUIC trap, consisting
of a pair of anti-Helmholtz coils and a third coil in perpendicular
orientation~\cite{Esslinger1998}. The adiabatic transport is realized by adding a
second anti-Helmholtz pair. A numerical simulation is used to
calculate optimized currents for all coils in order to create a
slowly changing trap and a smooth transport to the final position
while maintaining a magnetic field for proper spin orientation of
the sample. By these means, it is also possible to accelerate and
launch degenerate ensembles with high precision and reproducibility.
This capability is of great interest for its application in fountain
clocks~\cite{Wynands2005} and inertial sensors~\cite{Yver-Leduc2003}.

The transport mechanism presented in this paper is particularly
useful for applications where low heating rates and large atom
numbers are required. In particular, it may be used to load chip
traps or optical interferometers \cite{Dumke2002} with large atomic
samples. It may also be used to transport atoms to probe specific
position dependent quantities \cite{Gunther2005}. By cascading
the coil configuration used in this experiment, it will be possible to
cover much larger transport distances.

In our case, the transport
is used to load the atomic cloud into a dipole trap located at the
geometric center of the QUIC trap. Transporting the cloud to this
position enables us to use the coils of the QUIC trap to generate strong
homogeneous magnetic fields with small spatial inhomogeinity. Hence,
the applicability of the popular QUIC trap is drastically improved,
since it can be used to produce strong magnetic fields (around $1000~{\rm G}$)
at moderate currents. The use of such fields has recently become
important for the experimental manipulation of the scattering
properties of ultracold ensembles in the vicinity of Feshbach
resonances \cite{Inouye1998,Inouye2004,Ferlaino2006,Klempt2007}.

The paper is organized as follows. We give an overview of our
experimental setup in Sec.~\ref{sec:setup}. Details of our
implementation and results of the transport of quantum degenerate
gases are discussed in Sec.~\ref{sec:transport}. We conclude with an
outlook in Sec.~\ref{sec:summary}.

\section{Experimental Setup}
\label{sec:setup} 
\begin{figure}
\centering
\includegraphics*[width=\columnwidth]{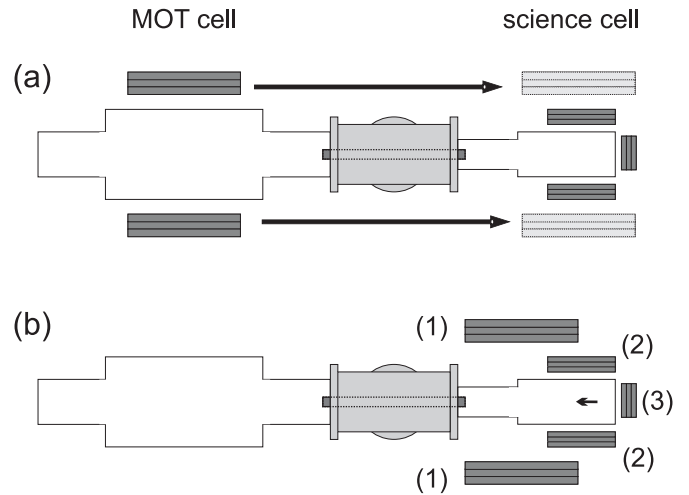}
\caption{Outline of the vacuum system and magnetic field coils. The
glass cell of the MOT is on the left hand side. The atoms are
transported to the science cell on the right hand side by a magnetic
field induced by a moving pair of coils as illustrated in~(a). A
differential pumping stage enables a lower pressure in the science
cell. The three coils of the QUIC trap around the science cell are
shown on the right hand side. The configuration~(b) is used for the
transport of the cold mixture to the geometric center of the QUIC
trap. The pair of transport coils~(1) is displaced from the
geometric axis defined by the main coils of the QUIC trap~(2),
opposite to the QUIC coil~(3).} \label{fig:vacuum}
\end{figure}

The apparatus used for the experiments described
here consists of two glass cells divided by a differential pumping
stage: a MOT cell, where the atomic clouds are collected initially,
and a science cell where experiments with ultracold atoms are
performed (see Fig.~\ref{fig:vacuum}). The MOT region, designed for
the collection of large clouds of K and Rb, has been described in
detail previously~\cite{Klempt2006}.

Atoms are transferred between these two regions of the experiment,
by transporting them in a movable magnetic quadrupole trap. This
transport mechanism is described in detail, since it is a key
element for further transport experiments with quantum degenerate
samples.

\subsection{Dual species MOT design}
The MOT is produced in a large glass cell with inner dimensions
$50~{\rm mm} \times 50~{\rm mm} \times 140~{\rm mm}$ at a pressure
of  $1~\times~10^{-9}~{\rm mbar}$. The cell allows for trapping
beams with a diameter of  $3~{\rm cm}$ to capture a large number of
atoms. Commercial rubidium dispensers and potassium dispensers
constructed according to Ref.~\cite{DeMarco1999a} are used to
provide vapors of $^{87}{\rm Rb}$ and $^{40}{\rm K}$. These
dispensers are located at a distance of $35~{\rm cm}$ from the MOT
cell and coat the surfaces of the chamber with rubidium and
potassium.

Two high power laser systems are necessary to provide the light for
magneto-optical trapping of the two atomic species. The cooling and
repumping light for the rubidium atoms is provided by two external
cavity diode lasers. Both beams are superposed and simultaneously
amplified by a tapered amplifier (TA) chip~\cite{Walpole1996}. A
further external cavity diode laser amplified by a TA provides
resonant light at the $^{39}{\rm K}$ D2 transition frequency.
This light is divided into two parts and one part is shifted
to the $^{40}{\rm K}$ cooling frequency by an acousto-optical
modulator (AOM) in double-pass configuration. The second part is
tuned resonant to the $^{40}{\rm K}$ repumping by an AOM in
quadruple-pass configuration and recombined with the cooling light.
After further amplification of the light for potassium with a second
TA, it is combined with the light for rubidium with a long pass
mirror. A single polarization maintaining fiber collects all four
frequencies for the cooling of potassium and rubidium. The use of a
single fiber greatly facilitates all further adjustments and the
optic setup is not more complicated than for a single species MOT.
We operate the MOT with a total power of $360~{\rm mW}$ for Rb and
$160~{\rm mW}$ for K.

The performance of this MOT is further improved by the use of
light-induced atom desorption (LIAD) at a wavelength of $395~{\rm
nm}$. Atoms that are adsorbed at the walls of a vacuum chamber are
desorbed by irradiation with weak incoherent light. This allows for
a temporary increase of the desired partial pressure. LIAD can thus
be used to load a rubidium MOT~\cite{Anderson2001} and to obtain
high loading efficiencies~\cite{Atutov2003}. Our recent
experiments~\cite{Klempt2006} have shown that LIAD is particularly
well suited as a switchable atom source, since the pressure decays
back to equilibrium after the desorption light is turned off. For
the experiments described here about $1~\times~10^8$ $^{40}{\rm K}$
and $5~\times~10^9$ $^{87}{\rm Rb}$ are trapped while the desorption
light is on, then they are held in the MOT without desorption light
while the pressure drops and finally, only magnetic fields are used
to confine them.

\subsection{Magnetic transport}
After the desorption light is switched off, the temperature of the
atoms is brought close to the recoil limit in an optical molasses
phase. In a second preparation step the atoms are optically pumped
to the fully stretched states $|f,m_f>=|2,2>$ for Rb and $|9/2,9/2>$
for~K. This allows to capture the atoms in a magnetic quadrupole
field induced by two MOT coils. These are mounted on a translation
stage. Just before it starts moving, the current of these coils is
ramped up from $14~{\rm A}$ to $28~{\rm A}$ in $300~{\rm ms}$, which
compresses the cold ensemble.

The coils are moved over a distance of $42~{\rm cm}$ to the science
cell at a pressure of $2~\times~10^{-11}~{\rm mbar}$ (see part~(a)
in Fig.~\ref{fig:vacuum}). Such transport systems for cold atoms
have previously been realized using moving
coils~\cite{Lewandowski2003} or sets of overlapping
coils~\cite{Greiner2001}, since these systems do not require a
second MOT in the science cell and provide far better optical access
in this region.

The magnetic confinement for the transport of cold thermal samples
is provided by the same coils which produce the small field
gradient for the MOT. These coils can produce a quadrupole field
with a gradient of up to $138~{\rm G}/{\rm cm}$. Each coil has
132~turns of $1~{\rm mm}\times 2.5~{\rm mm}$ copper wire. The coils
of $13~{\rm mm}$ thickness are separated by $74~{\rm mm}$ and have
an inner diameter of $45~{\rm mm}$. Their wires are fixed by epoxy
resin to avoid drifts in the performance. No active cooling is
needed.

This pair of coils is mounted on a translation stage (Parker, 404~XR
series), with a nominal position reproducibility of $5~\mu{\rm m}$.
However, the motor of the translation stage is switched off if it is
at rest to avoid rf noise between $1$ and $30~{\rm MHz}$, which
perturbs the rf evaporation of rubidium. The switching does not
effect the position. This experimental technique may be used to
transport atoms of two species together~\cite{Goldwin2004} or for
uniting cold clouds~\cite{Bertelsen2007}.

The translation stage is a reliable, maintenance free tool in our
experiments. Although the motion can be controlled in detail, we
have chosen a simple operation meth\-od. We allow for a maximal speed
of $10~{\rm m}/{\rm s}$, limit the acceleration to $1 ~{\rm m}/{\rm
s}^2$, and the jerk to $100~{\rm m}/{\rm s}^3$. With these settings,
the distance of $0.42~{\rm m}$ is covered in less than $1.3~{\rm
s}$. Similarly to other experiments~\cite{Theis2004,Greiner2001}, at
least one third of the particles reach the science cell. The losses
can be accounted to collisions with the background gas and to the
transport through the differential pumping tube between the two
glass cells. Moreover, we have not observed heating during the
transport and can conclude that this type of transport is well
suited for the transport of mixtures of different species.

The mechanical transport has been suggested for mixing of cold gases
with many components, which may be created in different MOT regions
and combined with moving coils~\cite{Bertelsen2007}. This may be
especially useful for combinations, which are more difficult to
combine in a MOT than K and Rb. Moreover, a chain of such traps has
been suggested as a step towards a continuously created
BEC~\cite{Lahaye2006}.

\subsection{Production of quantum degenerate gases}
The efficient transfer of cold atoms from the MOT region into a
harmonic trap in the science cell enables the production of quantum
degeneracy for both $^{87}{\rm Rb}$ and $^{40}{\rm K}$ by forced
rf-evaporation of rubidium. Therefore, the quadrupole field for the
transport is converted into a harmonic trapping potential, produced
by a magnetic trap in QUIC configuration [see Graph~(b) in
Fig.~\ref{fig:vacuum}].

To load the atoms into the QUIC trap, they are first transfered from
the transport coils into an even stronger quadrupole formed by the
main coils of the QUIC trap. Subsequently, the current through the
QUIC coil is ram\-ped up. Thus, the atoms are pulled towards this coil
and the trap is deformed such that effective evaporation to quantum
degeneracy is possible~\cite{Esslinger1998}. The $^{87}{\rm Rb}$
atoms are cooled by rf-evaporation until a Bose-Einstein condensate
with up to $1.5\times~10^6$ atoms at a temperature $T=460~{\rm nK}$
with a transition temperature $T_{\rm C}=580~{\rm nK}$ is reached.
The $^{40}{\rm K}$ atoms are cooled sympathetically with the Rb
atoms down to the same temperature reaching a quantum degenerate
Fermi gas with $1.3\times~10^6$ K atoms (with a Fermi temperature of
$T_{\rm F}=1530~{\rm nK}$).

The QUIC trap consists of a set of coils in anti-Helm\-holtz
configuration with $92$~turns with a separation of $30~{\rm mm}$ and
a third QUIC coil with $86$~turns [see (2) in Graph~(b) of
Fig.~\ref{fig:vacuum}], which is offset from the center of the
anti-Helmholtz pair by $40~{\rm mm}$ [see (3) in Graph~(b) of
Fig.~\ref{fig:vacuum}]. This coil configuration produces an offset
field of $1.4$~G and trapping frequencies for $^{87}{\rm Rb}$ of
$23~{\rm Hz}$ axially and $240~{\rm Hz}$ radially. In this
configuration the same current of $25~{\rm A}$ flows through all
three coils, yielding an offset field stability of $3~{\rm mG}$. All
experimental results presented in this paper were acquired by
releasing the atomic clouds from the magnetic confining potential
and taking resonant absorption images after ballistic expansion.

\section{Transport in a harmonic trap}
\label{sec:transport} Many recent experiments with cold ensembles
utilize homogeneous magnetic fields to manipulate the interaction of
cold atoms in the vicinity of Fesh\-bach
resonances~\cite{Weiner1999}. In our experiments the homogeneous
magnetic fields are created by the main coils of the QUIC
trap~\cite{Klempt2007}, a solution that reduces the experimental
complexity in the proximity of the cell. This allows us to profit
from the achieved high mechanical and current stability of the QUIC
trap.

\begin{figure}
\centering
\includegraphics*[width=\columnwidth]{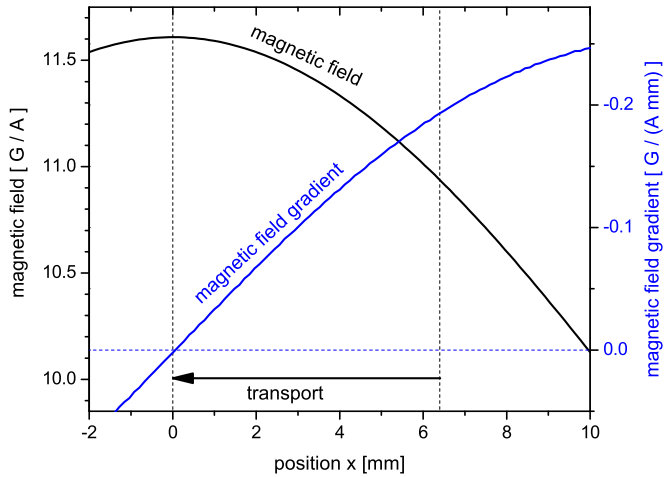}
\caption{Calculated magnetic field strength and its gradient
generated by the main coils of the QUIC trap operated in Helmholtz
configuration as a function of the radial position in the symmetry
plane. The transport to the geometric center of the QUIC trap
results in a smaller variation of the magnetic field strength over
the size of the size of the cloud.} \label{fig:noField}
\end{figure}

However, any displacement of the atoms from the symmetry axis leads
to a variation of the magnetic field over the width of the cloud. This
variation is much smaller in the geometric center of the pair of
main coils as shown in Fig.~\ref{fig:noField}. A transport of the
cold mixture to this region will therefore lead to a smaller
magnetic field spread and thus a better control of this crucial
parameter.

We have implemented such a transport of ultracold or quantum
degenerate samples over a distance of $\approx 6$~mm in a harmonic
trap. This transport is realized by extending the capabilities of a
magnetic trap in QUIC configuration~\cite{Esslinger1998} with an
additional coil pair in anti-Helmholtz configuration. In our case
that coil pair is identical to the one for the transport to the
science cell. Part~(b) of Fig.~\ref{fig:vacuum} illustrates the coil
configuration around the vacuum system for this transport.

The basic idea is to obtain a quadrupole field induced by the two
coil pairs (main and transport), which is shifted away from the one
induced by the main coils alone. For any position, the QUIC coil can
be used to convert the effective quadrupole field to a harmonic
trap.

We have simulated the magnetic fields to obtain optimal parameters
for the timing of all currents to realize the transport. The
position of the transport coils was chosen such that full optical
access to the cell is guaranteed. Precise control of the translation stage
enables us to position the atoms exactly in the center of the main coils.

For the transport of a quantum degenerate atomic ensemble in the
magnetic trap, one has to consider both the position of the potential
minimum and the offset field, which corresponds to the trapping
frequencies. In particular, the offset field should never drop to
zero during the transformation, otherwise spin flips quickly lead to
loss from the trap and destroy a quantum degenerate ensemble.
Another goal of our optimization strategy was to quickly reduce
the radial trapping frequency in order to minimize density dependent
losses and heating.

\begin{figure}
\centering
\includegraphics*[width=0.7\columnwidth]{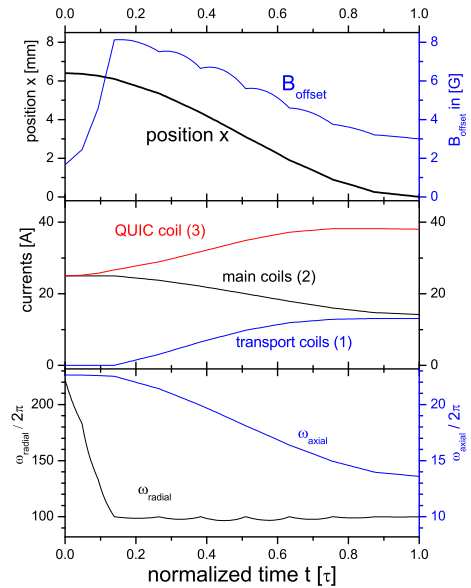}
\caption{Simulation of the transport from the QUIC trap to the
symmetry axis of the quadrupole coils. The top graph shows the
resulting positions and the offset magnetic field. The middle graph
illustrates the currents as obtained in an optimization with twelve
points. The wobbles for the magnetic field are due to the linear
interpolation between these points. The trap frequencies for
$^{87}{\rm Rb}$ are depicted in the bottom graph. Note that the trap
frequencies for $^{40}{\rm K}$ can be interfered directly using the
ratio of the masses since the product $m_f\,g_f=1$ for the
respective transported states is equal.} \label{fig:sim}
\end{figure}

An optimization algorithm was applied to find suitable currents
through the coils taking these issues into account. We aimed for the
following functional behavior of the position~$x$ with time~$t$
\begin{equation}
x\left(t\right) = D \frac{1}{2} \left[ \cos\left(\pi \frac{t}{\tau} \right) +1 \right]
\label{eq:move}
\end{equation}
for a total transfer from position~$D$ to the origin in the
time~$\tau$. Thus, the velocity of the trap changes steadily from
and to zero. The result of the simulation is shown in
Fig.~\ref{fig:sim}. To realize the transfer, the current of the
additional quadrupole coil pair and the current through the QUIC
coil are simultaneously increased while the current through the
anti-Helmholtz coil pair is decreased. The center of the harmonic
potential shifts towards the common center of the two quadrupole
coil pairs.

\begin{figure}
\centering
\includegraphics*[width=\columnwidth]{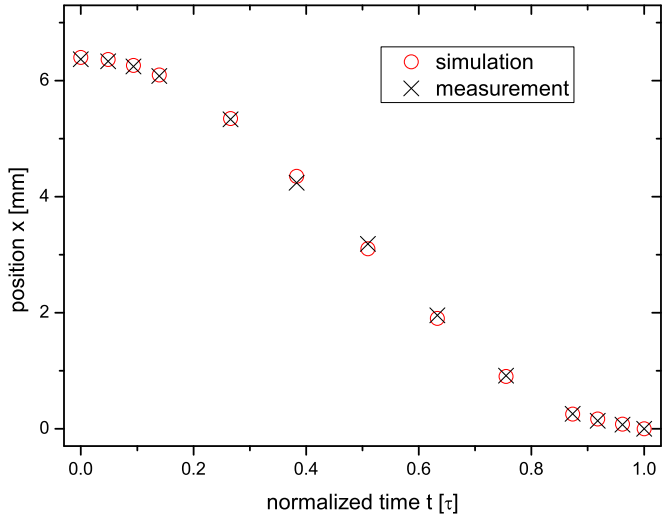}
\caption{Simulated position of the harmonic trap and measured
positions of a sample of cold rubidium.} \label{fig:xt}
\end{figure}

By controlling the currents through the coils, the transfer can in
principle be realized adiabatically. Since it is experimentally
necessary to implement the time dependence of the currents in linear
ramps we have limited our simulation to twelve linear ramps.

Based on this simulation, we have experimentally implemented the
transport and took absorption images after each of the applied
ramps. The atoms follow the position of the trap precisely, as
illustrated in Fig.~\ref{fig:xt}. The measurement confirms the
simulated transport of the ultracold mixture. It shows that we have
complete control over the position of the atoms by applying designed
currents through the coils to induce an adiabatically changing
magnetic field.

\begin{figure}
\centering
\includegraphics*[width=\columnwidth]{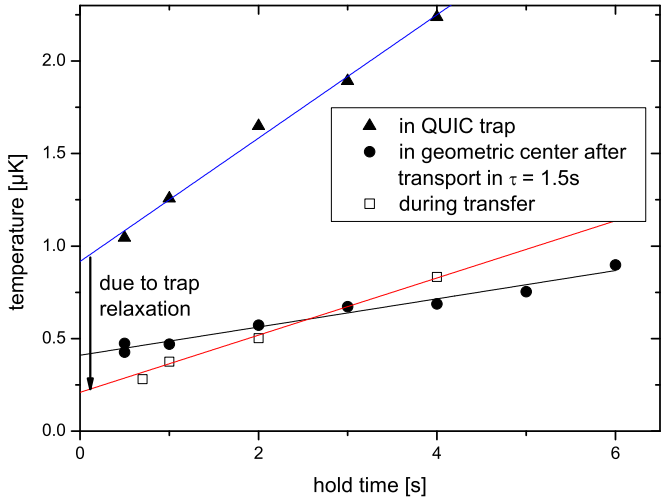}
\caption{Heating of a $^{87}{\rm Rb}$ ensemble just above $T_{\rm
C}$ in either a magnetic trap or while being transported. For the
later, the temperature in the magnetic trap at the end of the
transport was measured for different transport times~$\tau$.}
\label{fig:heat}
\end{figure}

Due to the lower trapping frequencies the heating rate of the sample
in the final trap is lower than in the initial QUIC trap ($75$ and
$330~{\rm nK}/{\rm s}$). The observed heating during the transfer is
between these two rates and thus no additional heating due to the
transport is present as shown in Fig.~\ref{fig:heat}. In fact, it is
possible to transport quantum degenerate gases without loss of the
degeneracy if the transport time~$\tau$ is shorter than $1~{\rm s}$.

\begin{figure}
\centering
\includegraphics*[width=0.7\columnwidth]{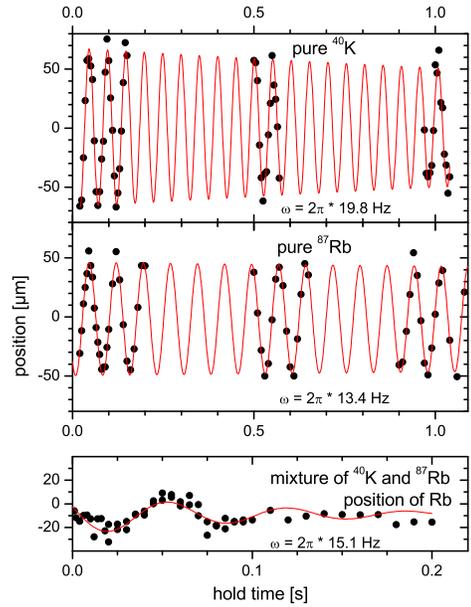}
\caption{Oscillations in the harmonic trap after the transfer from
the QUIC trap to the geometric center for pure samples and a mixture
of rubidium and potassium. Note the change in the time scale for the
mixture, in which the oscillation of rubidium is not only smaller
but also strongly damped by the potassium atoms.} \label{fig:oscill}
\end{figure}

Due to the chosen functional behavior [see Eq.~(\ref{eq:move})] in
particular short transport times $\tau < 1.5~{\rm s}$ lead to an
oscillation of the atomic clouds in the final magnetic trap (see
Fig.~\ref{fig:oscill}). In this case the trap is not changing
adiabatically anymore and the atoms act like a classical particle in
a harmonic trap. We have experimentally determined that twelve ramps
represent a good compromise between adiabaticity of the transport
and experimental complexity.

We obtain similar results for the transport of a heteronuclear
mixture. Due to the different mass, the $^{40}{\rm K}$ atoms
oscillate with a different frequency. We observed that the amplitude
of the oscillations of a cold cloud of a single species is
significantly reduced if Rb and K~atoms are transported together
(see Fig.~\ref{fig:oscill}). Similar effects in the hydrodynamic
regime were observed and studied in greater detail in
Ref.~\cite{Ferlaino2003}. For transport times above $1.5~{\rm s}$,
no significant oscillations can be observed and the twelve ramps of the
currents taken directly from the simulations need no fine
adjustments.

\begin{figure}
\centering
\includegraphics*[width=\columnwidth]{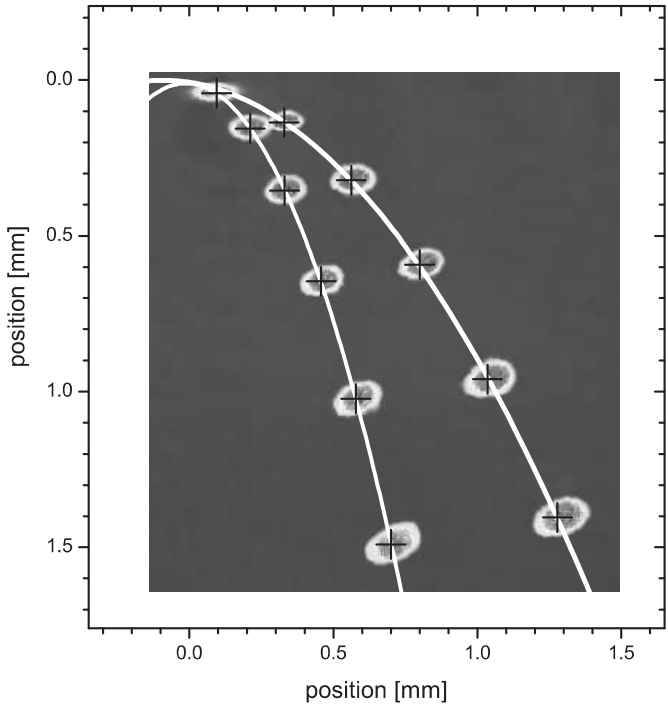}
\caption{Trajectories of a BEC launched with two different
horizontal speeds. Several absorption images are compiled into a
single image and overlay with two parabolas fitted to the obtained
positions of the clouds. From these fits, velocities of  $40~{\rm
mm}/{\rm s}$ and $80~{\rm mm}/{\rm s}$ can be interfered, which agree
with the calculated speed of the harmonic trap at the time when all
currents are rapidly switched off.} \label{fig:parabolar}
\end{figure}

In addition to the transfer mechanism our method allows us to
accelerate and launch ultracold ensembles by quickly switching off
all currents in the middle of the transfer.
Figure~\ref{fig:parabolar} shows BECs launched with horizontal
velocities of up to $80~{\rm mm}/{\rm s}$. This technique is an
alternative for launching cold ensembles with optical
lattices~\cite{Schmid2006} or detuned laser fields, which is usually
applied in fountain clocks~\cite{Wynands2005} or in inertial
sensors~\cite{Yver-Leduc2003,Muller2007}. Also, the observation time of quantum
degenerate gases can be doubled if the sample is launched against
gravity, provided that the time of flight is the limiting factor.

\section{Summary and Outlook}
\label{sec:summary} We have developed a transport mechanism for
quantum degenerate gases in a harmonic trapping potential and we
have demonstrated the simultaneous transport of quantum degenerate
bosonic and fermionic samples over a distance of up to $6~{\rm mm}$.
This mechanism may be cascaded to cover even larger distances and
thus enables magnetic transport experiments with large quantum
degenerate samples in macroscopic trap configurations. This concept
adds another powerful method to the toolbox of quantum atom optics
and will allow novel designs for interferometric sensors and clocks.

This transport mechanism enriches the possible applications of the
popular QUIC trap geometry significantly. It allows for a transport
of a quantum degenerate sample to the geometric center of the main
coil pair of the QUIC trap. This enables their use for the
production of large homogeneous fields and thus the investigation
and utilization of Fesh\-bach resonances, e.g., for the creation of
heteronuclear dimers or the tuning of the interaction of the two
trapped isotopes. Moreover, the transport facilitates the optical
access for additional beams, e.g., for the creation of optical
lattices or optical pumping such as photoassociation.

In the case of our experiment, the modification has reduced the
spread of a magnetic field of $500~{\rm G}$ from $240~{\rm mG}$ to
below $16~{\rm mG}$ for ensembles at $1~\mu{\rm K}$. This variation
of the magnetic field is no longer due to the inhomogeneity of the
field but due to residual current noise. Such well controlled
magnetic fields in combination with a dipole trap allow for the
precise control of the effective interaction strength and open the
pathway to studies of many particle physics such as the phase
separation between Fermions and Bosons or molecular physics such as
cold molecule production.

\section{Acknowledgments}
We acknowledge support from the Deutsche Forschungsgemeinschaft
(SFB~407 and Graduate College \emph{Quantum Interference and
Applications}).

\bibliographystyle{prsty}
\bibliography{all}
\end{document}